\documentclass[useAMS,usenatbib]{mn2e}

\usepackage{amsmath,amssymb,dsfont}
\usepackage{graphicx}
\usepackage{txfonts}
\usepackage[pdftex]{lscape}
\usepackage{rotating}

\title[Functional biases in GRB's spectral parameter correlations]{Functional biases in GRB's spectral parameter correlations}
\author[F. Massaro, S. Cutini, M. L. Conciatore and A. Tramacere]{F. Massaro$^{1}$\thanks{E-mail:
massaro@roma2.infn.it}, S. Cutini$^{2,3}$, M.L. Conciatore$^{2,4,5}$ and A.
Tramacere$^{5}$\footnotemark[1]\thanks{This file has been amended to
highlight the proper use of \LaTeXe\ code with the class file.
These changes are for illustrative purposes and do not reflect the
original paper by F. Massaro.}\\
$^{1}$Dipartimento di Fisica, Universit\`a di Roma Tor Vergata, Via della Ricerca scientifica 1 , I-00133 Roma, Italy\\
$^{2}$ASI Science Data Center, ESRIN, I-00044 Frascati, Italy\\
$^{3}$Dipartimento di Fisica, Universit\'a di Perugia, Viale A. Pascoli 1, I-06123 Perugia, Italy\\
$^{4}$INAF - Osservatorio Astronomico di Roma, Via Frascati, 33, 00040, Monteporzio Catone (Roma), Italy\\
$^{5}$Dipartimento di Fisica, Universit\`a di Roma La Sapienza, Piazzale A. Moro 2, I-00185 Roma, Italy.\\}

\begin{document}

\date{Accepted ...... Received ...... ; in original form ......}

\pagerange{\pageref{firstpage}--\pageref{lastpage}} \pubyear{2002}

\maketitle

\label{firstpage}

\begin{abstract}
Gamma Ray Bursts (GRBs) show evidence of different spectral shapes, light curves, duration, host galaxies and they explode within a wide redshift range. However, the most of them seems to follow very tight correlations among some observed quantities relating to their energetic. If true, these correlations have significant implications on burst physics, giving constraints on theoretical models. Moreover, several suggestions have been made to use these correlations in order to calibrate GRBs as standard candles and to constrain the cosmological parameters. \\
We investigate the cosmological relation between low energy $\alpha$ index in GRBs prompt spectra and the redshift $z$.
We present a statistical analysis of the relation between the total isotropic energy $E_{iso}$ and the peak energy $E_p$ (also known as Amati relation) in GRBs spectra searching for possible functional biases. \\
Possible implications on the $E_{iso}$ vs $E_p$ relation of the $\alpha$ vs $(1+z)$ correlation are evaluated.
We used MonteCarlo simulations and the boostrap method to evaluate how large are the effects of functional biases on the $E_{iso}$ vs $E_p$.
We show that high values of the linear correlation coefficent, up to about 0.8, in the $E_{iso}$ vs $E_p$ relation are  obtained for random generated samples of GRBs, confirming the relevance of functional biases. \\
Astrophysical consequences from $E_{iso}$ vs $E_p$ relation are then to be revised after a more accurate and
possibly bias free analysis.
\end{abstract}


\section{Introduction}
Gamma-ray Bursts (GRBs) are brief and intense flashes of high energy radiation emitted mostly in the $\gamma$-ray band. 
They are detected from wholly random directions in the sky at the rate of about once a day and typically last from a few milliseconds to several minutes. 
Within a few years, the BATSE experiment on board the NASA's Compton Gamma Ray Observatory satellite (Fishman et al. 1989) has recorded over 2700 GRB events with an isotropic distribution in the sky (Meegan et al. 1996).
However, although BATSE was very sensitive to high-energy photons, it could not discern the direction of a burst to better than a few degrees uncertainty, too large to pinpoint the location of individual explosions.

The real revolutionary step forward occurred in the 1997, thanks to the Italian-Dutch BeppoSAX satellite (Boella et al. 1997). This satellite was not as sensitive as BATSE to $\gamma$ rays, but its relative quick response of pointing system, coupled with good accuracy position information, permitted the first detection of an X-ray {\it afterglow}, the radiation emitted after the initial burst of $\gamma$-ray (Costa et al. 1997).  This discovery of afterglows made redshift measurements possible and confirmed that GRBs lie at cosmological distance ($0.0085$ (Galama et al. 1999) $<z<$ $6.29$ (Kawai et al. 2006)). 

It is well known that in the past years several correlations have been discovered linking various energies characterizing GRBs. 
All of them involve $E_p$, the energy peak of time integrated spectral energy distribution (SED). 
The first relation found links the rest frame isotropic energy $E_{iso}$ with $E_p$ (Amati et al. 2002, AM02 hereinafter, also known as 'Amati relation', see also Amati, 2006, AM06 hereinafter, for an updated version). 
This correlation is seen by several authors as an useful method to standardizing GRB energetics.

Subsequently, $E_p$ was found to be tightly correlated also with the collimation corrected energy $E_{\gamma}$ (Ghirlanda et al. 2004, GH04 hereinafter, also known as 'Ghirlanda relation'), and this relation is used to constrain 
cosmological parameters using GRBs as 'known' candles. 
The same relations can be expressed in terms of luminosity using spectra not time integrated: $L_{iso}\propto E_{p}^k$ (Yonetoku et al. 2004), $L_{\gamma}\propto E_{p}^k$ (Ghirlanda, Ghisellini \& Firmani 2006).

However, this kind of relations seems to contradict some observational evidences, as the large variety of light curves, spectra, redshifts, durations and host galaxies, leading us to suppose that the nature of GRB explosions is not unique.

Band et al. (2005) and  Nakar et al. (2005) tested the consistency of a large sample of BATSE GRBs (with unknown redshift) with the Amati and Ghirlanda relations. 
Their results suggest that they may be artifacts of selection effects, inferring that about half (Nakar et al. 2005) or even $\sim90\%$ (Band et al. 2005) of the whole GRB population cannot satisfy the correlation for any value of the redshift. 
However, these conclusions have been questioned by several other authors (Ghirlanda et al. 2005, Bosnjak et al. 2005, Pizzichini et al. 2005), that found instead that the peak energy and the fluences of BATSE GRBs with unknown redshift are fully consistent with the $E_{iso}~vs~E_{p}$ and the $E_{\gamma}~vs~E_{p}$ relations.

In AM02, and in AM06, $E_{iso}$ and $E_{p}$ are not evaluated directly by a fitting procedure but they are calculated using  analytic relations that include the same parameters.
These procedures could introduce functional biases in the spectral correlation quoted above. The influence of these functional biases has not been never considered in literature.

In this work we present a statistical analysis of the $E_{iso}$ vs $E_p$ relation, based on Monte Carlo and bootstrap simulations, searching for possible intrinsic correlation terms.
We previously study the cosmological relation between the redshift $z$ and the low energy spectral index $\alpha$ (see Sec.3), because a correlation between these spectral parameters can make more tight the $E_{iso}$ vs $E_p$ relation.
In Sect. 4 we evaluate functional and intrinsic correlation terms in the $E_{iso}$ vs $E_p$ relation, in order to estimate if the method used to derive it is statistically biased.

\section{The GRB Spectral Description}
GRBs have a non thermal spectrum that varies strongly from one burst to another. It is generally found that a simple power law does not fit well their spectra because of a continuos steepening toward the high energies.
An excellent phenomenological model was introduced by Band et al. (1993) to describe the prompt time-integrated GRBs spectra, composed by two power laws joined smoothly at a break energy $E_b$:
\begin{equation}
N(E)=\left\{
\begin{array} {lllllll}
A\left(\frac{E}{E_0}\right)^{\alpha}~exp\left(-\frac{E}{E_c}\right)&&& && E\leq E_b\\
B\left(\frac{E}{E_0}\right)^{\beta}& &&&&E\geq E_b\\ 
\end{array}
\right .
\end{equation}
where $N(E)$ is the number of photons per unit of area and energy, while $E_0$ is a reference energy usually fixed to the value of 100 keV. 
Under the continuity requirement for the function $N(E)$ and its first derivative, the break energy and normalization are given by:
\begin{equation}
E_b=(\alpha-\beta)~E_c 
\end{equation}
for typical values $\alpha\simeq -1.4$ and $\beta\simeq -2.4$, $E_{b}\approx E_{c}$, and
\begin{equation}
B=A~\left[\frac{(\alpha-\beta)E_c}{E_0}\right]^{\alpha-\beta}~e^{(\beta-\alpha)}
\end{equation}

There is no particular theoretical model that predicts this spectral
shape, however it provides good fits to most of the observed spectra
in terms of four parameters: the two photon indices $\alpha$ and
$\beta$, the exponential cut-off $E_c$ and the normalization constant
$A$, parameters directly estimated during the fitting procedure. 
The peak energy $E_p$ of the SED is related to the
spectral parameters by:
\begin{equation}
E_p=(\alpha+2)E_c < Ec
\end{equation}
being, typically, $-2<\alpha<-1$.

\section{The correlation between $\alpha$ and the redshift}
For the GRBs of BATSE catalogue a correlation between $\alpha$ and the redshift $z$ was found (Lloyd et al. 2000). 
It was interpreted as partially due to the dependence of the $\alpha$ on the spectral curvature around $E_p$ (Lloyd et al. 2000) and partially to the dependence of $E_p$ on the $\alpha$ index (Band et al. 1993).
A similar correlation was found in AM02 with a high logarithmic correlation coefficient $r_{log}=-0.83$ between the photon index $\alpha$ and $(1+z)$ (which corresponds to a linear correlation coefficient of the same quantities $r_{lin}=-0.77$). No clear explanation was given by the authors, who suggested that it could be a consequence of the $E_{iso}$ vs $E_p$ correlation. Even if a real satisfactory explanation for the $\alpha$ vs $(1+z)$ relation has not yet been found, we will show (Sect. 4), that it is relevant when studying the Amati and similar relations.

We first searched for its validity using the GRB samples in GH04 and in AM06, that is larger than the one of AM02. 
After excluding the bursts of AM02, because their parameters were evaluated in the GRB rest frame, we obtained two samples of 15 and 23  bursts, respectively. For both samples we found a significant correlation, with $r_{lin}=-0.61$ for GH04 bursts and $r_{lin}=-0.56$ for the AM06 sample, having the same trend of AM02. Logharitm correlation coefficents and p-chance probability are reported in Tab.1. 
We fitted the data in  AM02, GH04 and AM06 samples with a linear relation:
\begin{equation}
\alpha=m(1+z)+q
\end{equation}
The best fit values of parameters $m$ and $q$ are reported in Tab.1, while regression lines are plotted in Fig.1 for each GRB samples.

\begin{table*}
\caption{$\alpha$-z correlation}
\begin{flushleft}
\begin{tabular}{lclllllll}
\hline
Sample REF. & GRBs$(\#)$ & $z$ range & $r_{lin}$ & $P_{lin}$  & $r_{log}$  & $P_{log}$ &  $q$      &   $m$ \\  
\hline 
\noalign{\smallskip}
Amati et al., 2002         &  9 & 0.42 - 3.42 & 0.77 (0.10) & 0.006 & -0.83 (0.19) & 0.002 & -2.0688 & 0.32312 \\
Ghirlanda et al., 2004 & 15 & 0.0085 - 4.5 & 0.61(0.05) &  0.01 & -0.67 (0.15) & 0.004 &-1.466 & 0.14982 \\
Amati,          2006         & 23 & 0.0085 - 4.5 & 0.56 (0.04) & 0.04 & -0.63 (0.05) & 0.0008 &-1.3813 & 0.21449 \\
\noalign{\smallskip} 
\hline
\end{tabular}
\end{flushleft}
\end{table*}
 
A possible explanation of the $\alpha$ vs $(1+z)$ correlation could be due to a selection effect:
brighest GRBs correspond to flatter photon indicies $\alpha$. 
The correlation between $\alpha$ and the redshift implies that we are loosing faint GRBs at high $z$ values. 
In  Fig. 1 the GRBs number decreasing at high redshift is shown.\\
In the next section we show how this correlation affects the $E_{iso}$ vs $E_p$.
\begin{figure}
\includegraphics[height=9cm,width=8cm,angle=0]{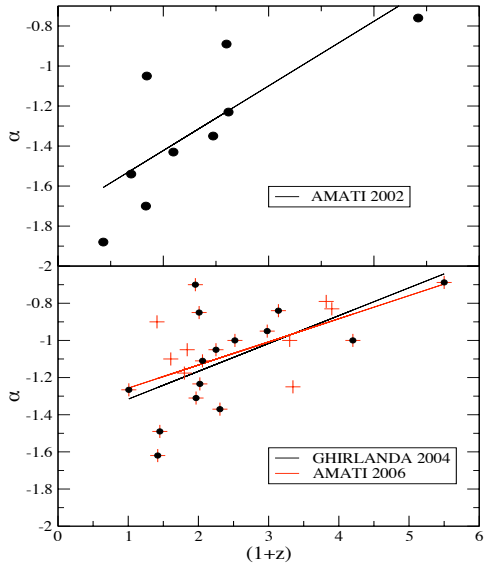}
\caption{The $\alpha$ vs $(1+z)$ relation. In the upper panel the blueshifted GRBs (black filled circles) of the AM02 sample are reported, while, in the lower panel,
there are GH04(black filled circles) and AM06(red crosses) bursts.In both panels the regression lines are plotted.}
\end{figure}

\section{The $E_{iso}$ vs $E_p$ relation}

It is clearly known that when two or more uncorrelated and independent
parameters are used to build up other variables, a correlation term
will arise between these new variables due to the functional relations
used in the calculations. So, in searching any kind of correlations
between these new variables it is necessary to take into account of
the possible biases due to the method adopted to estimated them.

AM02 reported the analysis of 12 GRBs with known redshifts observed by
BeppoSAX. For 9 blue-shifted spectra, using the Band model, the
spectral parameters ($\alpha$, $\beta$, $E_c$ and $A$) are estimated
in each GRB rest frame. The authors found a very tight correlation
($r_{lin}=0.96$) between $E_p$, defined by Eq.4 and:
\begin{equation}
E_{iso} = \frac{4\pi D_L^2}{(1+z)^2} \int_{1 keV}^{10^4 keV}E N(E;\alpha,\beta,E_c,A) dE
\end{equation}
the total (isotropic) energy emitted in the GRB rest frame. Both
quantities depend on the two parameter $\alpha$ and $E_c$ and it is
worth to be noted that $E_{iso}$ is a function of $z$, that is related
to $\alpha$ as discussed in the previous section.

In our analysis a flat Friedmann-Robertson-Walker Universe is assumed,
with $H_0=65$ km/(s Mpc), $\Omega_{M}=0.3$ and $\Omega_{\Lambda}=0.7$
as reported in AM02.  The luminosity distance $D_L$ was calculate
using the formula (Hogg 1999, Carroll et al. 1992):
\begin{equation}
D_L = (1+z) \frac{c}{H_0} \int_{0}^{z}\frac{dz}{\sqrt{\Omega_M (1+z)^3+\Omega_{\Lambda}}},
\end{equation}
that was also successfully compared with other cosmological
calculators (N. Wright and R. Priddey).  Our code was previously
tested using as input data the parameters reported in Tab.2 of AM02
and finding the same results for $E_{iso}$ and $E_p$.

In a recent work (Amati et al. 2006 and references therein), the
$E_{iso}$ vs $E_p$ correlation is evaluated using the following
relations that do not require bluesfhited GRBs spectra:
\begin{equation}
E_p=(\alpha+2)(1+z)E_c
\end{equation}
\begin{equation}
E_{iso} = \frac{4\pi D_L^2}{(1+z)} \int_{1 keV/(1+z)}^{10^4 keV/(1+z)}E N(E;\alpha,\beta,E_c,A) dE.
\end{equation}
We also tested functional biases in the method performed with Eq.8 and
Eq.9.

\subsection{Monte Carlo simulations: the $E_{iso}$ vs $E_p$ correlation}
Our investigation of a possible functional bias was made by building
up a numerical code that, after assigned the five spectral parameter
$z$, $\alpha$, $\beta$, $E_c$, $A$, calculates $E_{iso}$ and $E_p$
using Eq.4 and Eq.7.

To clarify the effect of redshift on the correlation, we initially
fixed its value at $z=1$, while the other spectral parameters
$\alpha$, $\beta$, $E_c$, $A$ were randomly, and so uncorrelated,
generated with an uniform distributions within the intervals of AM02
[-0.7; -1.9], [-2.1; -2.7], [340; 840] keV, [0.07; 2.27]
ph/(cm$^2$keV), respectively. Different values of $z$ were tested,
observing that the correlation do not depend on its choice.  Our
analysis was also performed with ranges of spectral parameters wider
than the observed ones, with consistent results.  $E_{iso}$ and $E_p$
values were calculated for each set of spectral parameters, iterating
this procedure for 45 GRBs, a factor of five larger than those
analysed in AM02.  Values of $E_p<90$ keV were discarded to be in
agreement with AM02.

Finally, the correlation coefficients $r_{lin}$, $r_{log}$ and the
non-parametric Spearman correlation coefficent $r_{spr}$, with their
standard deviations were estimated for this sample. This procedure
was repeated 1000 times to study the distribution of the correlation
coefficients.  The entire procedure have been also tested using 100
GRBs in each samle with consistent results with the previous analysis.

In Fig.2 (upper panel) the distribution of $r_{lin}$ is represented,
with a mean value of $<r_{lin}>=0.48$ and a standard deviation of
$\sigma_{lin}=0.13$.  The results obtained using the logarithmic
coefficient have a mean value $<r_{log}>=0.26$ and
$\sigma_{log}=0.14$, while the mean Spearman correlation coefficent is
$<r_{spr}>=0.33$ with a standard deviation $\sigma_{spr}=0.10$.  The
mean chance probability for the linear correlation coefficent is 0.03
and the probability to find a spurious correlation coefficent is not
negligible, being a relatively large number of linear correlation
coefficents higher than 0.6.

Sakamoto et al. (2006) shows for a sample of 32 GRBs that there 
is a correlation between the energy peak and the total fluence with $r_{log}=0.58$
Our simulations at fixed redshift, corresponding to the relation between these quantities, 
show that we found a correlation coefficient 
practically coincident with the Sakamoto et al. results.
Again this support the relevance of biases introduced by functional relation.    

To evaluate the cosmological influence due to the redshift on the
$E_{iso}$ vs $E_p$ correlation, we also generated the $z$ values
randomly. To taking into account the correlation relating to $\alpha$
and $z$, as discussed in the previous section, redshift values were
generated with an uniform distribution for each GRB in the range
[0.42; 3.42], while $\alpha$ indices were evaluated using the formula:

\begin{equation}
\alpha=m(1+z)+q+R
\end{equation}
The values of $m$ and $q$ are evaluated using AM02 data (see values reported in Tab.1), while $R$ is a random number within the range [-0.5,0.5]. 
\begin{figure}
\includegraphics[height=9cm,width=8cm,angle=0]{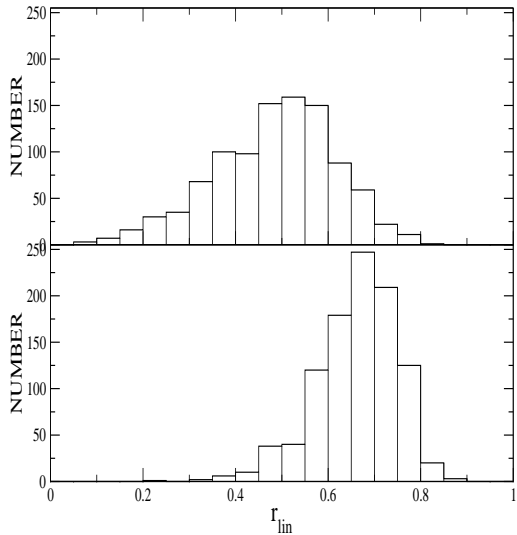}
\caption{The distribution of 1000 linear correlation coefficents evaluated with fixed redshift (upper panel) or taking into account of the intrinsic correlation between $\alpha$ and $z$ (lower panel)
using Eq.4 and Eq.5 for the Monte carlo simulations.}
\end{figure}
\begin{figure}
\includegraphics[height=9cm,width=8cm,angle=0]{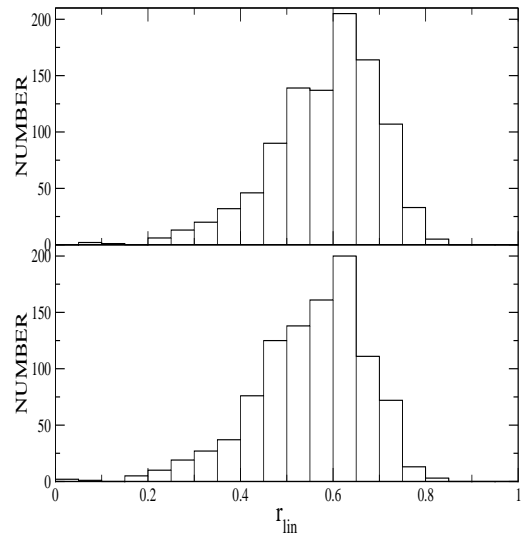}
\caption{The distribution of 1000 linear correlation coefficents evaluated using Eq.8 and Eq.9 instead of Eq.4 and Eq.5 and taking into account of the intrinsic correlation between $\alpha$ and $z$ 
in GH04 and AM06 for the Monte carlo simulations.}
\end{figure}
The introduction of $R$ allows us to have a correlation coefficient
$<r_{lin}>=-0.70$ between $\alpha$ and $(1+z)$ consistent with the
observed value (AM02).  The $\beta,~E_c,~A$ parameters were generated
as in the previous analysis.

By iterating this procedure to build 1000 correlation coefficients, we
found a value of $<r_{lin}>=0.66$ with $\sigma_{lin}=0.09$ (Fig.2,
lower panel) with a mean value of the chance probability of 3.7
$\times$ 10$^{-4}$ and a value of $<r_{log}>=0.54$ with
$\sigma_{log}=0.12$ and $<r_{spr}>=0.55$ and $\sigma_{spr}=0.10$.

Following GH04 and AM06 relations taking into account of the
intrinsic correlation between $\alpha$ and $(1+z)$ and of the different range of redshift (see Tab. 1),
we have found an even more higher correlation coefficents, as reported in Tab.2.  This
is beacuse $E_{iso}$ and $E_{p}$ in Eq.8 and Eq.9 depend both on the
$\alpha$ index, the high energy cut-off $E_c$ (Eq.5), and by the
redshift $z$.  The distributions of linear correlation coefficents are
reported in Fig.3.

It is worth to be noted that if you decrease the number of simulated
GRBs to build a single correlation coefficient, for example from 45
down to 9 as the AM02 sample, the distribution of $r_{lin}$ becomes
more spread and the frequency of $r_{lin}$ values close to $1$ is not
negligible ($r_{lin}>0.7$ for about half values).

In Tab.2 all the correlation coefficents values, calculated taking
into account of the $\alpha$ vs $(1+z)$ correlation, and their
p-chance probabilities are summarized.

In the correlations analysis could be relevant the ranges of the quantities involved 
in the evaluation of spectral parameters.
To verify our results we performed the same simulations using parameter ranges wider than the observed ones, for example in the case of the normalization we used an interval 
spanning up to five order of magnitude, and we obtain the same results.   

\begin{table}
\caption{Correlation coefficents}
\begin{flushleft}
\begin{tabular}{llllllll}
\hline
Relations &  $r_{lin}$ & $\sigma_{lin}$ & $P_{lin}$ &  $r_{log}$ & $\sigma_{log}$ &  $r_{spr}$ & $\sigma_{spr}$  \\  
\hline 
\noalign{\smallskip}
AM02 & 0.66 & 0.09 & 3.7 $\times$ 10$^{-4}$ & 0.54 & 0.12 & 0.55 & 0.10 \\
GH04 & 0.58 & 0.11 & 4.7 $\times$ 10$^{-3}$ & 0.51 & 0.12 & 0.37 & 0.13 \\
AM06 & 0.55 & 0.12 & 7.9 $\times$ 10$^{-3}$ & 0.48 & 0.12 & 0.35 & 0.13 \\        
\noalign{\smallskip} 
\hline
\end{tabular}
\end{flushleft}
\end{table}

\subsection{Monte Carlo simulations: the $k$ exponent in the $E_{p}\propto E_{iso}^{k}$ relation}
We also checked our results evaluating the exponnt in the  $E_{iso}$ vs $E_p$ relation.
The $k$ exponent in the $E_{p}\propto E_{iso}^{k}$ relation evaluated in AM02 with 9 GRBs has a value of $0.52$.

In Fig. 4 we report our distribution of the $k$ exponents in the
relation $E_{p}\propto E_{iso}^{k}$, with a mean value of $<k>=0.26$
and a dispersion of $\sigma_{k}=0.05$.

We observe that in AM02 GRBs sample there are two peculiar GRBs: GRB
990123 having $E_p$ outside of the observed energy range and GRB
010222 for which there is only a lower limit on $E_p$ based on an
assumption on the $\beta$ index.  Excluding from the AM02 sample these
two GRBs, the obtained value for the $k$ exponent is $0.36$, in
agreement with the analysis performed by Amati et al. (2003, 2006),
and it is consistent with our results.  It is worth to be noted also
that values of $k$ exponents in Amati et al. (2006) are in the range
[0.35; 0.57] and seem to depend by the sample used to evaluate the
$E_{iso}$ vs $E_p$ relation.
\begin{figure}
\includegraphics[height=9cm,width=8cm,angle=0]{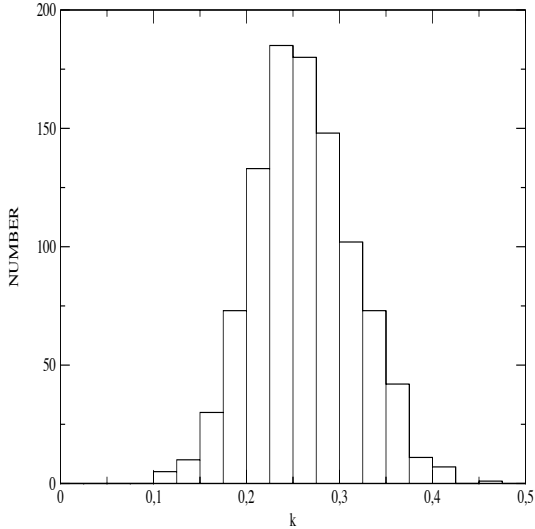}
\caption{The distribution of $k$ exponents in the simulated GRBs of the $E_{iso}$ vs $E_p$ relation.}
\end{figure}

\subsection{Boostrap method}
\begin{figure}
\includegraphics[height=9cm,width=8cm,angle=0]{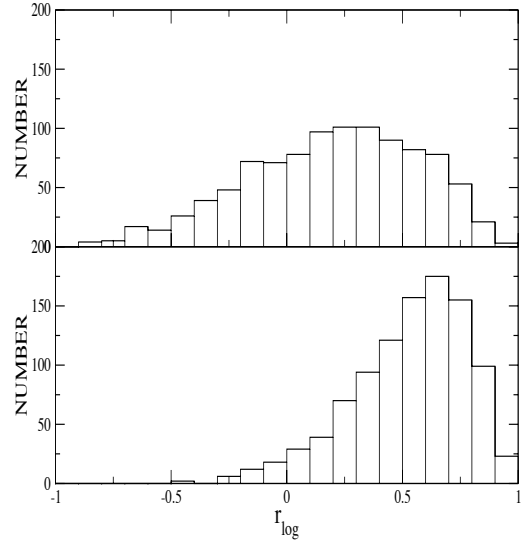}
\caption{The distribution of 1000 logarithm correlation coefficents for the boostrap simulations (upper panel) and taking into account of the intrinsic correlation between $\alpha$ and $z$ (lower panel).}
\end{figure}
\begin{figure}
\includegraphics[height=9cm,width=8cm,angle=0]{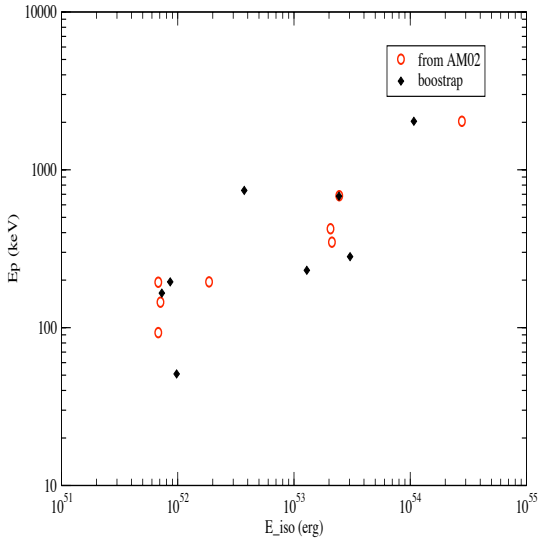}
\caption{Black filled points represent 8 GRBs simulated with the
boostrap method, in comparison with those in AM02 sample indicated
with red circles.}
\end{figure}
Our results were also checked performing also a bootstrap method, a
techinique frequently adopted in the statistical analysis of
correlations (Manly, 1997), that consists in the random association of
the parameter's values between the elements in the same sample.

To perform this method we randomly selected the spectral parameters
$z,~\beta,~E_{c},~A$ of eight GRBs of the sample used in AM02 (see
Tab.1 and Tab.2 in AM02). We excluded GRB010222 because in AM02 only a
lower limit of the value of the $E_p$ is reported. We found values of
the logarithm correlation coefficent $<r_{log}>=0.23$ with
$\sigma_{log}=0.44$.

To take into account the correlation between $\alpha$ and $z$, their
values were selected belong to the same GRB.  Following this method,
we found an higher value of $<r_{log}>=0.52$ with a standard deviation
of $\sigma_{log}=0.24$.

The results of the bootstrap method (Fig.5) are consistent with the
Monte Carlo simulations.  In Fig. 6 the coverage of our bootstrap
simulated GRBs in comparison with observed ones (AM02) in the
$E_{iso}$ vs $E_p$ plane is shown.

\section{Discussion}
In this paper a statistical analysis of the GRBs parameters
correlations $\alpha$ vs $(1+z)$ and $E_{iso}$ vs $E_p$ is presented.
 We studied the $\alpha$ vs $(1+z)$ correlation observing that it is
present not only in the AM02 GRB sample but also in those analysed by
GH04. As stated by different authors (Lloyd et al. 2000, Band et al. 1993),
there is no clear explanation for it, but we suggested that it could be a direct observational
consequence of a selection effects: faint high redshift GRBs are under
the detectability threshold.  As a consequence of this, the decrease
of high redshift GRBs introduce a correlation between $\alpha$ and
$z$.
This correlation can have important consequences when we attempt to
use GRBs as standard candles and, moreover, it introduces a bias in
searching relations between quantities builded using $\alpha$ and $z$.
 
We investigated how the presence of functional correlation terms
affects the $E_{iso}$ vs $E_p$ relation. As builded in AM02, $E_{iso}$
and $E_p$ depend on a common set of spectral parameters. 
We performed several Monte Carlo simulations , initially considering
only the intrinsic correlation term at a fixed redshift. We found that
the $E_{iso}$ vs $E_p$ relation is biased, with a not negligible
probability to reproduce correlation coefficents higher than 0.5.
When we introduced in the generation of simulated sample also the
$\alpha$ vs $(1+z)$ correlation, the mean value of the correlation
coefficent increase up to $0.64$. Higher values were obtained using
Eq.8 and Eq.9 instead of Eq.4 and Eq.5. A boostrap method was also
applied to the AM02 sample. The good agreement with the previous
analysis proves that our results are independent on the simulation
code.

An already discussed (e.g. Band et al. 2005) relevant subject is the
contribution from selection effects in making more tight the $E_{iso}$
vs $E_p$ relation with respect to our simulations.  It is possible
that instrumental sensivity limits would produce forbidden regions in
the plane $E_{iso}, E_p$. As an example, a combination of the spectral
parameters in the GRB rest frame could make a GRB with a very low
luminosity and therefore below the detectability threshold. We also
stress that the SED energy peak is an averaged quantity, because
during a single GRB that shows more bumps in its lightcurve we expect
that $E_p$ changes, making difficult the link of this variable to
physical quantities.\\

In the literature appear several other tight correlations concerning
the energetics and the spectral distributions of GRBs (e.g
$E_{\gamma}~vs~E_p$ (GH04), $L~vs~E_p$ (Yonetoku et al. 2004)), some
of them are used to constrain cosmological parameters, using GRBs as
'known' candles. All of them are based on the $E_{iso}$ vs $E_p$
relation and this implies that they are affected by functional biases.
In the GH04 relation, for example, the collimation corrected energy is
computed considering the jet opening angle, that is in turn derived
from $E_{iso}$ and $z$. \\

In any case, we do not exclude that a physical relation between the
$E_{iso}$ and $E_p$ can really exist, but it can be safely established
only after removing all the biases.

A good solution could be to use spectral laws explicitly written in
terms of the quantities for which a correlation is searched (as made,
for example, in Tramecere, Massaro \& Cavaliere (2007) for the
spectrum of the near HBL object Mkn 421).

A further step could also be to work with homogeneous samples of GRBs
in terms of some relevant characteristics (e.g. time evolution,
spectral shape, ...) to verify that the correlation is actually
followed by them.

\section*{Acknowledgments}

     We thank E. Massaro for suggestments in using the theory of correlation and for a critical revision of the
     manuscript, P. Giommi, G. Tosti and G. Consolini for useful discussions and comments.
     We thank also F. Berrilli and S. Giordano for useful suggestions in the pratical use of Monte Carlo simulations.  
     F. Massaro acknowledges a grant by the Italian Space Agency (ASI) for the AGILE Space Mission.

\bsp

\label{lastpage}

\end{document}